\documentclass[aps,prb,twocolumn,superscriptaddress,showpacs,floatfix]{revtex4}

\usepackage{graphicx}

\newcommand{\bra}{\langle}
\newcommand{\ket}{\rangle}
\newcommand{\mr}{{\mathbf{r}}}
\newcommand{\mR}{{\mathbf{R}}}
\newcommand{\mN}{{\mathbf{0}}}
\newcommand{\mk}{{\mathbf{k}}}

\begin{document}

\title{An ab-initio many-body method for electronic structure calculations
  of solids.\\
  II. Unscreened Hartree-Fock treatment for the 3d systems Fe, Co, Ni and Cu}

\author{I. Schnell}
\affiliation{Theoretical Division,
  Los Alamos National Laboratory, Los Alamos, New Mexico 87545}
\affiliation{Department of Physics, University of Bremen, P.O.Box 330 440,
  D-28334 Bremen, Germany}

\author{G. Czycholl}
\affiliation{Department of Physics, University of Bremen, P.O.Box 330 440,
  D-28334 Bremen, Germany}

\author{R.C. Albers}
\affiliation{Theoretical Division,
  Los Alamos National Laboratory, Los Alamos, New Mexico 87545}

\date{\today}

\begin{abstract}
The ab-initio many-body method suggested in the preceding paper is
applied to the 3d transition metals Fe, Co, Ni, and Cu. We use a
linearized muffin-tin orbital calculation to determine Bloch functions
for the Hartree one-particle Hamiltonian, and from these obtain
maximally localized Wannier functions. Within this Wannier basis all
relevant one-particle and two-particle Coulomb matrix elements are
calculated. The resulting second-quantized many-body Hamiltonian with
ab-initio parameters is studied within the simplest many-body
approximation, namely the unscreened, selfconsistent, Hartree-Fock
approximation (HFA). We present these HFA results, which we believe
are the first to have been done for crystalline 3d
transition metals, and compare them with those obtained from the
standard local (spin) density approximation (LSDA) within density
functional theory (DFT).  Although the d-bands sit considerably lower
within HFA than within L(S)DA, the exchange splitting and magnetic
moments for ferromagnetic Fe, Co, and Ni are only slightly larger in
HFA than what is obtained experimentally or within LSDA. The HFA total
energies are lower than the corresponding L(S)DA calculations.
\end{abstract}


\maketitle

\section{\label{sec:Introduction}Introduction}

In the preceding paper\cite{Schnell2003-1} we suggested a new method
for ab-initio electronic-structure calculations of solids. The main
steps of this procedure are:
\begin{enumerate}
\item Perform a conventional, self-consistent, band-structure
  calculation for an effective one-particle Hamiltonian, namely, the
  Hartree Hamiltonian, to obtain a suitable basis set of Bloch
  functions.
\item By taking into account only a finite number $J$ of bands one
  chooses a truncated one-particle Hilbert space. The
  Marzari-Vanderbilt\cite{MV97} algorithm is then used to construct a
  maximally localized set of Wannier functions, which span the same
  truncated one-particle Hilbert space.
\item All one-particle (tight-binding) and two-particle (Coulomb)
  matrix elements of the Hamiltonian within this Wannier function basis
  are calculated.
\item The resulting electronic many-body Hamiltonian in second
  quantization with parameters determined from first principles is
  studied within standard many-body approximations for lattice electron
  systems.
\end{enumerate}
In this paper we apply this scheme to the 3d transition metals Fe, Co,
Ni, and Cu. We use the ``linear muffin-tin orbital'' (LMTO) method
within the ``atomic-sphere approximation'' (ASA)\cite{Skriver} to
perform the band-structure calculation for the Hartree Hamiltonian in
first quantization. The direct Coulomb matrix elements of the
maximally localized Wannier basis are rather large, about 20 eV in
magnitude. We then use the simplest possible many body approximation,
the Hartree-Fock approximation (HFA), to study the second-quantized
multiple-band Hamiltonian.  Our results are compared with those
obtained from a standard LSDA calculation.\cite{DrGross90,Eschrig96}
Although the 3d-bands and the 4s-band overlap in the L(S)DA
approximation, our unscreened HFA calculations give 3d-bands that lie
considerably lower (between 10 and 20 eV) than the 4s-band.  The HFA
correctly predicts ferromagnetism for the ferromagnetic metals Fe, Co,
and Ni and no magnetism for Cu, but with a much larger exchange
splitting between majority and minority 3d bands than obtained within
LSDA and with a slightly larger magnetic moment per site than obtained
experimentally or within LSDA.  On the other hand, the total energy is
lower in HFA than in LSDA.
The LSDA results for metals are probably more reliable than our new
HFA results, which lack important screening and correlation effects.
In order for our method to go beyond LSDA we would need
to use better many body methods than the (unscreened) HFA, which
should be possible within our scheme.  

Our purpose in presenting the HFA results is to demonstrate that the
new ab-initio many-body method that we have proposed is feasible and
can be applied to practical calculations of materials.  In addition, because
HFA is the assumed standard starting point before adding complex many-body
correlations, and because it is the simplest many-body approximation
(or the best mean-field one-particle approximation), it
is useful to know what the HFA predicts for the 3d transition metals,
which have been so heavily studied by other techniques.  Comparisons
of HFA results with experiments, other (higher order) theories, or
established standard methods such as the L(S)DA should
demonstrate the effects of correlation on electronic properties in
d-electron systems.

To the best of our knowledge we do not know of any published HFA
results (band structure, density of states, magnetism, magnetic
moment, total energy, etc.) for the 3d ferromagnets Fe, Co and
Ni, unless it was implicitly applied to these materials for schemes
like the local ansatz\cite{Stollhoff}, where HFA results serve as an
input to higher order calculations .
The is not surprising since the HFA has, from very early on, been
viewed as a poor approximation
for metals. For example, when applied to
the homogeneous electron gas (as the simplest model of an infinite
metallic system), the HFA has well-known Fermi edge
singularities\cite{Mahan,Pisani}. These lead, in particular, to a
vanishing density of states (DOS) at the Fermi energy, which is, of
course, unphysical. This unphysical feature usually prevails in actual
HFA-calculations for real metals\cite{Monkhorst79}, though sometimes
this singularity is hard to see in actual
HFA-results\cite{DovesiPisani82}.  In our calculations the
non-locality is handled through the calculation of expectation values
(matrix elements of the density matrix), which makes HF calculations
as easy as Hartree calculations. Furthermore, because of our localized
Wannier basis, we only keep on-site and a few inter-site Coulomb and
exchange matrix elements.  Hence our calculations have an effective
short-ranged Coulomb interaction.  Although longer-range Coulomb
matrix elements are small in our calculations, which is why we
truncate them, it is possible that if all of them were kept to
infinite distances that they could add up to give Fermi edge
singularities (which are due to the long-ranged nature of the bare
Coulomb interaction) and other standard anomalies.  Correlation or
screening would quickly kill these effects.

The paper is organized as follows. In Section \ref{sec:LMTOHartree} we
describe the LMTO-Hartree calculations and the Hartree-results for the band
structure and density of states, respectively. Section
\ref{sec:Wannierfunctions} describes some results obtained for the maximally
localized Wannier functions obtained within the Marzari-Vanderbilt
algorithm\cite{MV97}, in particular their localization properties. Results
for the matrix elements, in particular the direct Coulomb and exchange
matrix elements are given in Section \ref{sec:matrixel}; we also
compare these results with calculations of the Slater integrals. 
The application of the
(unscreened) HFA to the multiband many-body Hamiltonian in second
quantization is the subject of Section \ref{sec:HartreeFock}. For an
interpretation of the results we compare the numerical HFA results obtained 
for the crystal with previous
atomic HFA results and with numerical and analytical results for a simplified 
local (atomic or zero band width) model in Section \ref{sec:atomicHFA}. 
A comparison with the more standard LSDA-results follows in
Section \ref{sec:LSDA},  before the paper closes with a short discussion.

\section{\label{sec:LMTOHartree}LMTO Hartree calculation}

For the four materials of interest (Fe, Co, Ni, Cu) 
we performed a selfconsistent Hartree band-structure calculation.
Besides the nuclear charge we used the (experimentally) known results for 
the lattice structure (bcc for Fe, fcc otherwise; Co should actually
be hexagonal) and for the lattice constant as input.
For the band-structure calculation we used the LMTO
method\cite{Skriver,OKA75} within the atomic sphere approximation (ASA). 
The radius of the (overlapping) muffin-tin spheres (the Wigner-Seitz
radius $S$) is determined by the condition that the sphere volume
equals the volume of the unit cell. This yields the following values for the
Wigner-Seitz radius: $S=2.662 a_0$ for Fe, $S=2.621 a_0$ for Co, $S=2.602 a_0$
for Ni and $S=2.669 a_0$ for Cu (Ref.~\onlinecite{Skriver}).
Within the muffin-tin spheres the
potential and wave functions are expanded in spherical harmonics with
a cutoff $l_{\rm max}=2$, i.e., s, p, and d-orbitals are included.

Within the LMTO-ASA the eigenfunctions,
i.e. the Bloch wave functions (inside a muffin-tin sphere),
are given in terms of the
solution to the radial Schr{\"o}dinger equation $\phi_{\nu l}(r)$ to
some fixed energies $E_{\nu l}$ and its energy
derivative $\dot\phi_{\nu l}(r)$:
\begin{equation} \label{eq:lmto-wave}
  \Psi_{n {\bf k}} ({\bf r}) = \sum_L \left(
    \phi_{\nu l}(r) A_L^{n {\bf k}} + \dot \phi_{\nu l}(r) B_L^{n {\bf k}}
  \right)
  Y_L( \hat {\bf r} ) ~~,
\end{equation}
where the $Y_L \equiv Y_l^m$ denote the (complex) spherical harmonics and
$n$ is the band index. 

\begin{figure}
\includegraphics[scale=0.45]{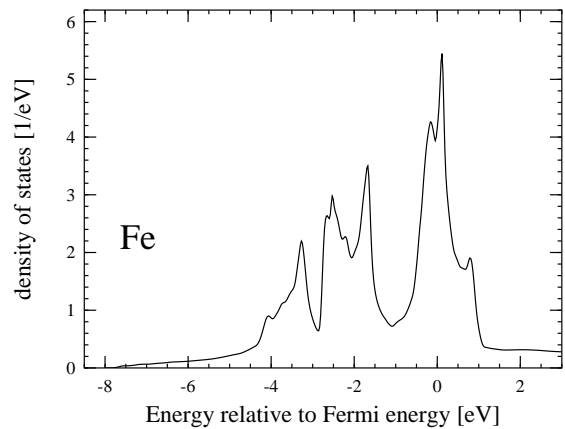}
\caption{\label{fig:feha} Total density of states (of both spin directions)
  around the Fermi level (chosen as zero of the energy axis
  $E_F=0$) obtained from the selfconsistent Hartree calculation for Fe.}
\end{figure}

\begin{figure}
\includegraphics[scale=0.45]{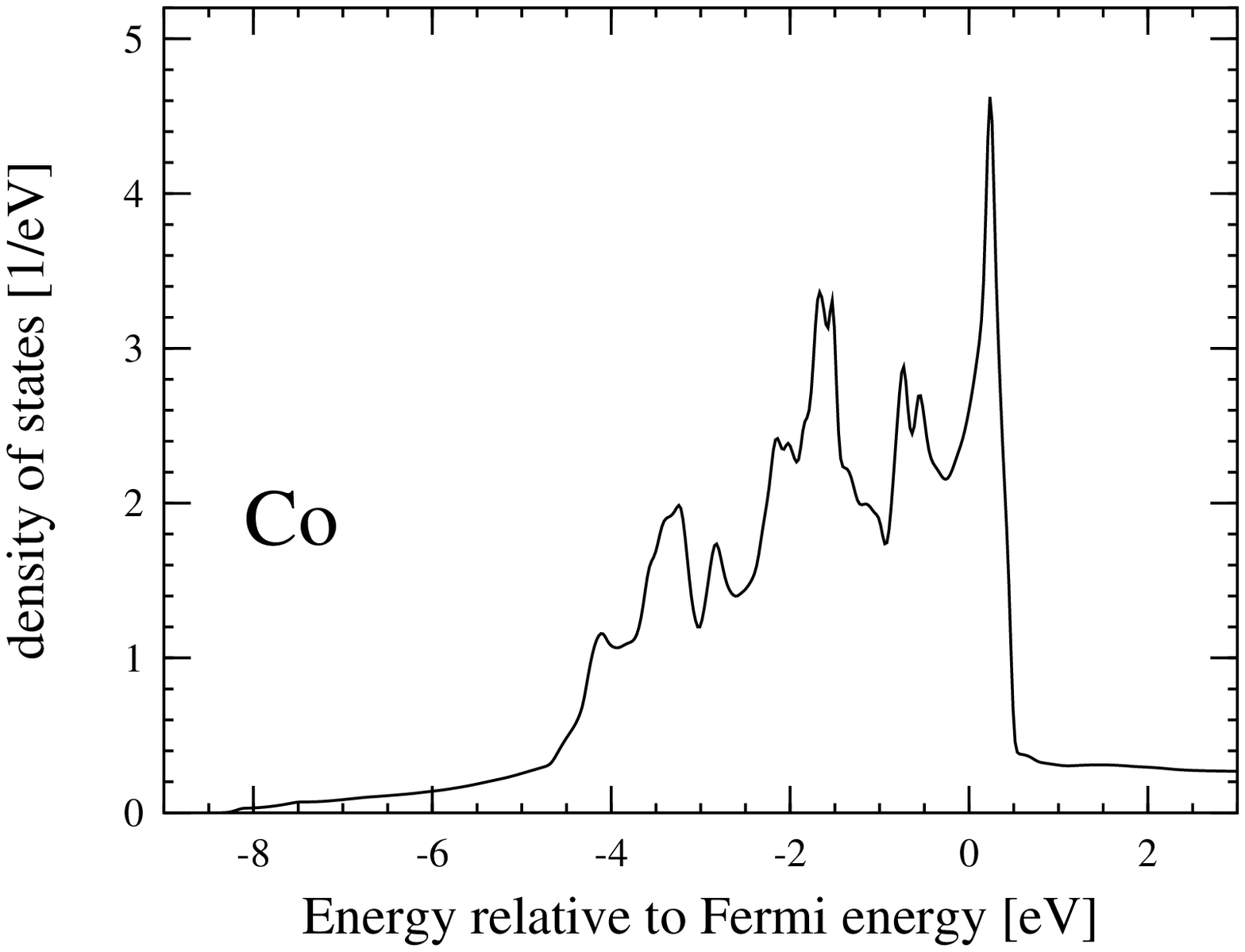}
\caption{\label{fig:coha} The same as in Fig. \ref{fig:feha} for Co.}
\end{figure}

\begin{figure}
\includegraphics[scale=0.45]{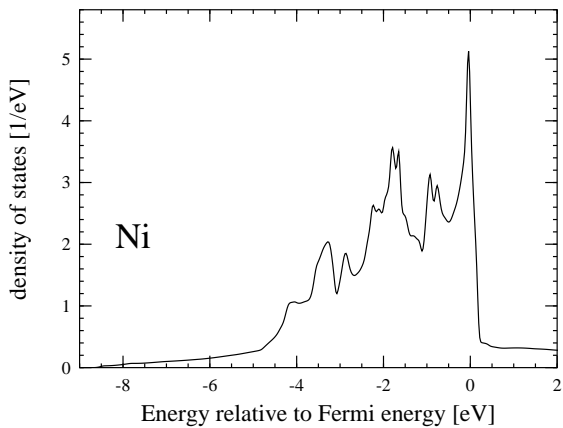}
\caption{\label{fig:niha} The same as in Fig. \ref{fig:feha} for Ni.}
\end{figure}

\begin{figure}
\includegraphics[scale=0.45]{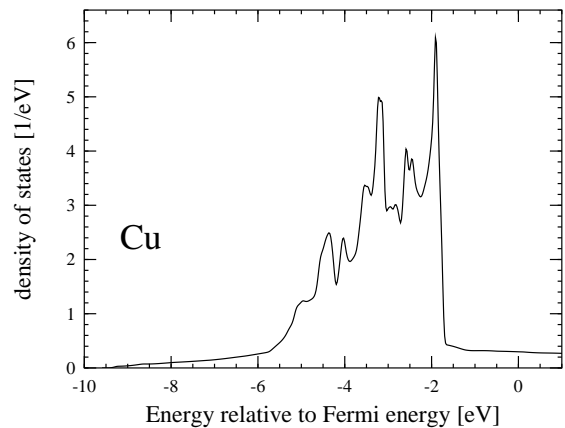}
\caption{\label{fig:cuha} The same as in Fig. \ref{fig:feha} for Cu.}
\end{figure}

Figs. \ref{fig:feha} - \ref{fig:cuha} show the nonmagnetic
(spin-degenerate) Hartree results for the density of states (DOS). One
observes that the five narrow 3d-bands are similar in energy and
therefore hybridize with the 4sp-bands. For Fe, Co, and Ni the Fermi level
$E_F$ falls in the midst of the partially filled 3d-bands, and hence there is a
large DOS at the Fermi level. For Cu $E_F$ is above the filled 3d bands,
where the states have more 4sp-like character and the DOS is small
(free electron like).
 
The Bloch functions obtained from this Hartree calculation are used as the
basis of a truncated Hilbert space. In the band calculation, there
is already a natural truncation due to the spherical harmonics expansion
cutoff $l_{max}=2$, which limits the calculation to 9 bands for one atom
per unit cell. The connection between the Bloch functions in the
different muffin-tin spheres is given by the standard Bloch relation
\begin{equation} \label{eq:p2}
  \psi_{n\mk}(\mr+\mR) = e^{i\mk\mR} \psi_{n\mk}(\mr) ~~.
\end{equation}
Hence the Bloch function in a single muffin-tin sphere determines the
function for the whole crystal. However,  this
simple relation holds for Bloch functions only. Any other function
(to be represented as linear combination of Bloch functions) can,
nevertheless, be decomposed into its contributions within the different
muffin-tins according to
\begin{equation} \label{eq:p3}
  \Phi_\alpha(\mr) = \sum_i \Phi_\alpha(\mR_i;\mr-\mR_i)
\end{equation}
where $\Phi_\alpha(\mR;\mr)=0$ for $|\mr|>S$.

\section{\label{sec:Wannierfunctions}Maximally localized Wannier functions}

We now turn to the construction and description of the Wannier functions and
will present results for their main properties for the four materials under
investigation. Because
\begin{equation}
  w_{\mR n}(\mr) = w_{\mN n}(\mr-\mR) ~~,
\end{equation}
it is sufficient to investigate the Wannier functions only for one
arbitrarily chosen lattice vector $\mN$: $w_n(\mr) \equiv w_{\mN n}(\mr)$.
This Wannier function has contributions not only in the muffin-tin sphere
around $\mN$ but also in other muffin-tin spheres, and we can use the
decomposition (\ref{eq:p3}) into the contributions from the different
muffin-tin spheres
\begin{equation} \label{eq:Wanndecomp1}
  w_n(\mr) = \sum_{\mR} w_n(\mR;\mr-\mR) ~~.
\end{equation}
After application of the Marzari-Vanderbilt algorithm, which is briefly 
summarized in the preceding paper\cite{Schnell2003-1}
and in more detail in Ref. \onlinecite{MV97}, 
the new set of bands that are used to calculate the
maximally localized Wannier functions can no longer be classified by pure
angular momentum quantum numbers. The Wannier functions are rather
admixtures having different $L$-contributions (3d, 4s, 4p etc.).
But, since the original Bloch functions from which
the Wannier functions are constructed
were given in terms of a spherical harmonics expansion, the new Wannier
functions (and their contribution in each individual muffin-tin sphere) 
can also be decomposed into these spherical harmonics contributions
\begin{equation} \label{eq:wann-wave}
  w_n(\mR;\mr) = \sum_L \left\{
    \phi_{\nu l}(r) A_L^{\mR n} + \dot \phi_{\nu l}(r) B_L^{\mR n}
  \right\}
  Y_L( \hat \mr ) ~~.
\end{equation}
One can then calculate the weight of the contributions to the Wannier
function (centered at $\mN$) within the different muffin-tin spheres
\begin{equation} \label{eq:wnR}
  \bra w_n | w_n \ket_\mR \equiv \int_\mR d^3\mr |w_n(\mr)|^2
  = \int_\mN d^3\mr |w_n(\mR;\mr)|^2 ~~,
\end{equation}
and one can also decompose this into the different $l$-contributions
according to:
\begin{equation} \label{eq:wnR_eval}
  \bra w_n | w_n \ket_\mR = \sum_l
  \underbrace{
    \sum_{m=-l}^l
    \left\{
      |A_{l m}^{\mR n}|^2 + \bra \dot\phi_{\nu l}^2 \ket |B_{ l m}^{\mR n}|^2
    \right\}
  }_{ \equiv ~ C_l^{\mR n} }
\end{equation}
For the 3d-system iron these quantities are tabulated in Table
\ref{tab:fe_wf}. The first line is the weight $\bra w_n|w_n \ket_{\mN}$
in the center muffin-tin. Between 88 and 98\%
of the total weight of the Wannier functions is to be found already within
the center muffin-tin; this shows how well localized our Wannier functions
are with the lowest five functions having values of more than 95\%. 
Rows 2--4
in this table indicate the different $l$-contribution or $l$-character of the
Wannier functions. One sees that the optimally localized Wannier
functions are not pure within their $l$-character, but the 
lowest five Wannier functions 
(0-4) still have mainly $l=2$ (3d) character. 
Higher band-index states (which are slightly less well localized according to
row 1) are admixtures that have mainly $l=1$ (4p) character (about 50 \%),
but also a considerable amount of $l=0$ (4s) and $l=2$ (3d) character.
Corresponding results for the other 3d-systems Co, Ni, and Cu are similar
and, therefore, not repeated here.

\begin{table}
  \renewcommand{\arraystretch}{1.2}
  \begin{tabular}{c|ccccccccc}
    $n$ & 0 & 1 & 2 & 3 & 4 & 5 & 6 & 7  \\
  \hline
$\sum_l C_l^{\mN n}$ & .9761 & .9765 & .9596 & .9800 & .9773 & .8754 & .8731 & .8763  \\
$\sum_\mR C_{l=0}^{\mR n}$ & .0019 & .0018 & .0081 & .0019 & .0017 & .2224 & .2381 & .2265  \\
$\sum_\mR C_{l=1}^{\mR n}$ & .0955 & .0726 & .1797 & .0611 & .0728 & .5480 & .5509 & .5347  \\
$\sum_\mR C_{l=2}^{\mR n}$ & .9026 & .9256 & .8121 & .9370 & .9255 & .2295 & .2110 & .2388  \\
  \end{tabular}
  \caption{\label{tab:fe_wf} Some properties of the lowest eight maximally
    localized Wannier functions of Fe.}
\end{table}

\section{\label{sec:matrixel}One particle and Coulomb matrix elements}

From the optimally localized Wannier functions we calculate the
one-particle matrix elements
\begin{equation}
  t_{12} = \int d^3\mr~ w_1^*(\mr)\left(-\frac{\hbar^2}{2m}\nabla^2 +
  V(\mr)\right)w_2(\mr)
\end{equation}
and the Coulomb matrix elements of the Hamiltonian
\begin{equation} \label{eq:W1234}
  W_{12,34} = \int d^3\mr~ d^3\mr'~ w^*_1(\mr) ~ w^*_2(\mr') ~~
  \frac{e^2}{|\mr-\mr'|} ~~ w_3(\mr') ~ w_4(\mr) ~~.
\end{equation}
Here we use the abbreviated notation 1 to mean
$\mR_1n_1$ and 2 to mean for $\mR_2n_2$, etc.
We used two different numerical algorithms to calculate these Coulomb matrix
elements, namely the FFT-algorithm briefly described in the previous paper
and a spherical expansion algorithm described in some detail 
in Ref. \onlinecite{SCA02}.
The latter method makes use of the fact that (in each muffin-tin sphere)
the Wannier functions are explicitly given as linear combinations of
products of spherical harmonics and a radial wave function.
The expansion
\begin{equation} \label{eq:p37}
  \frac{1}{|\mr-\mr'|} = \sum_{k=0}^\infty
  \frac{4\pi}{2k+1} ~ \frac{r_<^k}{r_>^{k+1}} \sum_{m=-k}^k
  Y_K^*(\hat\mr') ~ Y_K(\hat\mr)
\end{equation}
($K=\{k,m\}$)
makes it possible to express the on-site Coulomb integrals as
one-dimensional integrals over products of the radial functions and
the (tabulated) Gaunt coefficients. The results obtained by this algorithm
and by the independent FFT-algorithm agree within the numerical errors 
(of at most 1\%).

\begin{table}
  \renewcommand{\arraystretch}{1.2}
  \begin{tabular}{c|ccccccccc}
     $U_{nm}$ & 0 & 1 & 2 & 3 & 4 & 5 & 6 & 7 & 8 \\
  \hline
0 & 22.42 & 20.90 & 20.10 & 20.96 & 20.86 & 14.16 & 13.32 & 13.96 & 13.50 \\
1 & 20.90 & 23.04 & 19.95 & 21.55 & 21.53 & 14.07 & 13.54 & 13.58 & 14.15 \\
2 & 20.10 & 19.95 & 20.77 & 20.05 & 19.83 & 12.95 & 13.46 & 13.37 & 13.22 \\
3 & 20.96 & 21.55 & 20.05 & 23.27 & 21.67 & 13.46 & 14.05 & 13.98 & 13.98 \\
4 & 20.86 & 21.53 & 19.83 & 21.67 & 22.99 & 13.71 & 13.28 & 14.25 & 14.12 \\
5 & 14.16 & 14.07 & 12.95 & 13.46 & 13.71 & 13.67 &  9.45 &  9.58 &  9.64 \\
6 & 13.32 & 13.54 & 13.46 & 14.05 & 13.28 &  9.45 & 13.52 &  9.27 &  9.50 \\
7 & 13.96 & 13.58 & 13.37 & 13.98 & 14.25 &  9.58 &  9.27 & 13.75 &  9.65 \\
8 & 13.50 & 14.15 & 13.22 & 13.98 & 14.12 &  9.64 &  9.50 &  9.65 & 13.81 \\
  \end{tabular}
  \vspace{5mm}
  \quad
  \begin{tabular}{c|ccccccccc}
     $J_{nm}$ & 0 & 1 & 2 & 3 & 4 & 5 & 6 & 7 & 8 \\
  \hline
0 & 22.42 &  0.84 &  0.61 &  0.75 &  0.99 &  0.86 &  0.73 &  0.81 &  0.42 \\
1 &  0.84 & 23.04 &  0.77 &  0.88 &  0.84 &  0.70 &  0.51 &  0.48 &  0.86 \\
2 &  0.61 &  0.77 & 20.77 &  0.88 &  0.70 &  0.96 &  0.93 &  0.92 &  0.60 \\
3 &  0.75 &  0.88 &  0.88 & 23.27 &  0.82 &  0.33 &  0.78 &  0.64 &  0.69 \\
4 &  0.99 &  0.84 &  0.70 &  0.82 & 22.99 &  0.52 &  0.46 &  0.75 &  0.83 \\
5 &  0.86 &  0.70 &  0.96 &  0.33 &  0.52 & 13.67 &  0.58 &  0.56 &  0.57 \\
6 &  0.73 &  0.51 &  0.93 &  0.78 &  0.46 &  0.58 & 13.52 &  0.45 &  0.56 \\
7 &  0.81 &  0.48 &  0.92 &  0.64 &  0.75 &  0.56 &  0.45 & 13.75 &  0.55 \\
8 &  0.42 &  0.86 &  0.60 &  0.69 &  0.83 &  0.57 &  0.56 &  0.55 & 13.81 \\
  \end{tabular}
  \caption{\label{tab:fe_UJ} On-site direct and exchange
    Coulomb matrix elements between Wannier functions for Fe.
    All energies are eV's.}
\end{table}

Results for the on-site direct and exchange Coulomb
matrix elements between the optimally localized Wannier functions are
given in Table \ref{tab:fe_UJ}  for iron (Fe). 
The direct Coulomb integrals $U_{nm} = W_{nm,mn}$ between the Wannier states
with the lowest five
band indices ($n,m \in \{0,\ldots,4\}$), 
which according to the discussion and table
in the preceding section have mainly 3d-character, are rather large, up to 
23 eV for Fe. Within the 3d-like bands the 
interband direct Coulomb matrix elements are of the same magnitude as the
intraband matrix elements. The matrix elements between 3d-states and
4sp-states are considerably smaller, of the magnitude of 13 - 14 eV. For
electrons in 4sp-states ($n,m \in \{5,\ldots,9\}$) the direct intraband
Coulomb matrix elements are again of the order of 13 - 14 eV, but the
interband matrix elements are slightly smaller, about 9 eV.
The exchange matrix elements $J_{nm} = W_{nm,nm}$ are always 
much smaller, usually
less than 1 eV (for $n \neq m$). Again the corresponding 
results for the other 3d-systems
investigated (Co, Ni and Cu) are very similar.

\begin{table}
  \begin{tabular}{c|cccc}
   & $U$ & $J$ & $t_{NN}$ & $t_{NNN}$ \\
  \hline
  Fe & 21.1 & .81 & .59 & .24 \\
  Co & 22.6 & .87 & .55 & .10 \\
  Ni & 22.6 & .88 & .75 & .11 \\
  Cu & 24.5 & .94 & .80 & .12 \\
  \end{tabular}
  \caption{\label{tab:UJtaverage} Averaged on-site Coulomb, exchange,
    nearest neighbor and next nearest neighbor hopping matrix elements
    for the 4 3d-systems; energies are in eV.}
\end{table}

For the 5 states with predominant 3d-character we have calculated the averages
of the on-site direct and exchange Coulomb matrix elements
\begin{eqnarray} \label{eq:Fk_UJ}
  U &\equiv& \frac{1}{25} \sum_{mm'} W_{mm'm'm} \\ 
  J &\equiv& \frac{1}{20} \sum_{m\ne m'} W_{mm'mm'} ~~,
\end{eqnarray}
as well as the averages of the absolute values of the nearest neighbor (NN)
and next nearest neighbor (NNN) hopping matrix elements
\begin{equation}
  t_{NN(N)} \equiv \frac{1}{25} \sum_{n,m} |t_{\mR nm}| ~~.
\end{equation}
The results obtained thereby for the 4 transition metals under consideration
are shown in Table \ref{tab:UJtaverage}. The U-values vary between 21 eV for
Fe and 25 eV for Cu, the J-values are smaller than 1 eV and the hopping matrix
elements are of the magnitude 0.5 -- 0.7 eV for nearest-neighbor (NN) and
0.1 -- 0.2 eV for next-nearest-neighbor
(NNN), and further on decrease with increasing distance.

\begin{table}
  \begin{tabular}{l|ccc}
   & $F^0$ & $F^2$ & $F^4$ \\
  \hline
  Fe (crystal)                      & 21.62 &  9.61 & 5.91 \\
  Fe (atom [\onlinecite{Watson59}]) & 23.76 & 10.96 & 6.81 \\
  \hline
  Co (crystal)                      & 23.18 & 10.31 & 6.34 \\
  Co (atom [\onlinecite{Watson59}]) & 25.15 & 11.58 & 7.20 \\
  \hline
  Ni (crystal)                      & 24.69 & 11.00 & 6.77 \\
  Ni (atom [\onlinecite{Watson59}]) & 26.53 & 12.20 & 7.58 \\
  \hline
  Cu (crystal)                      & 26.27 & 11.72 & 7.23 \\
  Cu (atom [\onlinecite{Watson59}]) & 27.90 & 12.82 & 7.96 \\
  \end{tabular}
  \caption{\label{tab:F^k} Slater integrals $F^k$ (in eV) for the 3d-systems
    Fe, Co, Ni, Cu as obtained by our calculations and within an earlier
    atomic calculation\cite{Watson59}.}
\end{table}

We have also evaluated the Slater integrals\cite{S29}:
\begin{equation} \label{eq:Fk_def}
  F^k \equiv e^2 \int dr~  r^2 \int dr'~ r'^2 ~
  |R_{l=2}(r)|^2 ~ \frac{r_<^k}{r_>^{k+1}} ~ |R_{l=2}(r')|^2 ~~,
\end{equation}
where $R_{l=2}(r)$ is a radial (atomic) d-wave function (obtained by
solving the Schr{\"o}dinger equation for a radial symmetric potential, for
instance). Note that only the three integrals $F^0$, $F^2$ and $F^4$
are required to determine all the Coulomb $d$-matrix elements. 
Using the radial d-wave function obtained from the 
Hartree calculation we obtain the following values for the Slater integrals 
of the four
3d-systems: $F^0=$ 21.62 eV for Fe, 23.18 eV for
Co, 24.69 eV for Ni, and 26.27 eV for Cu. This means, the
Slater integrals $F^0$ are rather good estimates of our (averaged) 
Coulomb matrix elements. These values are also in agreement with older results
obtained in calculations for 3d-atoms\cite{Watson59}. In Table \ref{tab:F^k} we
show our $F^k$-values for the four 3d-crystals and compare them with
corresponding atomic calculations from Ref. \onlinecite{Watson59}. Obviously,
there is fairly good agreement between these atomic and our results.

\section{\label{sec:HartreeFock}Unscreened Hartree-Fock approximation}

After we have determined the matrix elements within our restricted basis set
of 9 maximally localized Wannier functions (per site and spin), we have a
Hamiltonian in second quantization of the form
\begin{equation} \label{eq:secondquant}
  H = \sum_{12\sigma} t_{12} c_{1\sigma}^{\dagger} c_{2\sigma} +
  \frac{1}{2} \sum_{1234\sigma\sigma'}
  W_{12,34} c_{1\sigma}^{\dagger}c_{2\sigma'}^{\dagger}c_{3\sigma'}c_{4\sigma}
\end{equation}
for which all the matrix elements are known from first principles. The
simplest approximation one can now apply is the HFA, which replaces the
many-body Hamiltonian by the effective one-particle Hamiltonian
\begin{eqnarray} \label{eq:HFHamilton}
  H_{HF} &=& \sum_{12\sigma} \left(t_{12} + \Sigma_{12,\sigma}^{HF}\right)
  c_{1\sigma}^\dagger c_{2\sigma} \\
\mbox{with } ~~ \Sigma_{12,\sigma}^{HF} &=& \Sigma_{12}^{Hart}
  + \Sigma_{12,\sigma}^{Fock}
\\ \nonumber
  &=& \sum_{34\sigma'} \left[ W_{13,42} -
  \delta_{\sigma\sigma'}W_{31,42}\right] \bra c_{3\sigma'}^{\dagger}
  c_{4\sigma'} \ket ~~.
\end{eqnarray}
Here the expectation values $\bra c_{1\sigma}^{\dagger}c_{2\sigma} \ket$
have to be determined selfconsistently for the HF Hamiltonian
(\ref{eq:HFHamilton}). Note that the Fock (exchange) term is
spin ($\sigma$) dependent and may, therefore, give rise to magnetic
solutions. 

\begin{figure}
\includegraphics[scale=0.42]{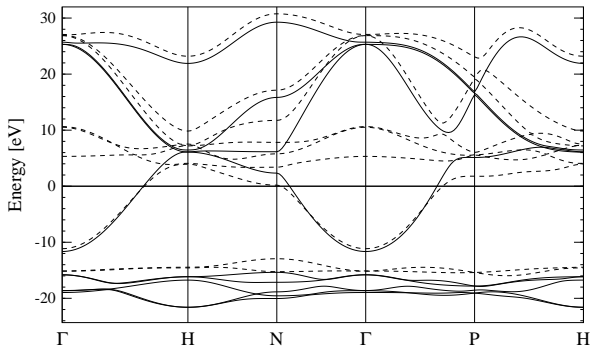}
\caption{\label{fig:fehfband} (Unscreened) Hartree-Fock band-structure
  of Fe; the full line shows the majority (spin up),
  the dashed line the minority spin component.}
\end{figure}

\begin{figure}
\includegraphics[scale=0.45]{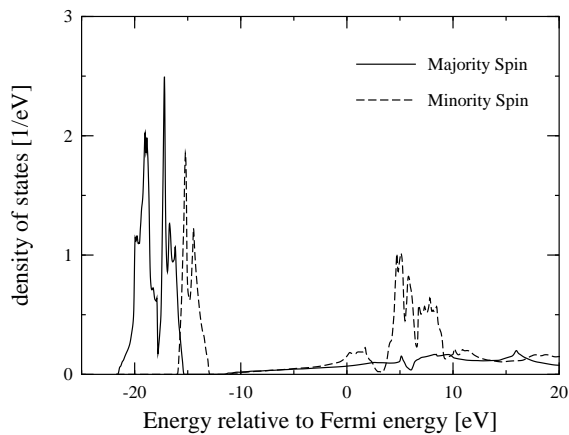}
\caption{\label{fig:fehfdos} Density of states (per spin direction) for
  Fe within the (unscreened) Hartree-Fock approximation; the full line shows
  the majority (spin up), the dashed line the minority spin contribution.}
\end{figure}

\begin{figure}
\includegraphics[scale=0.42]{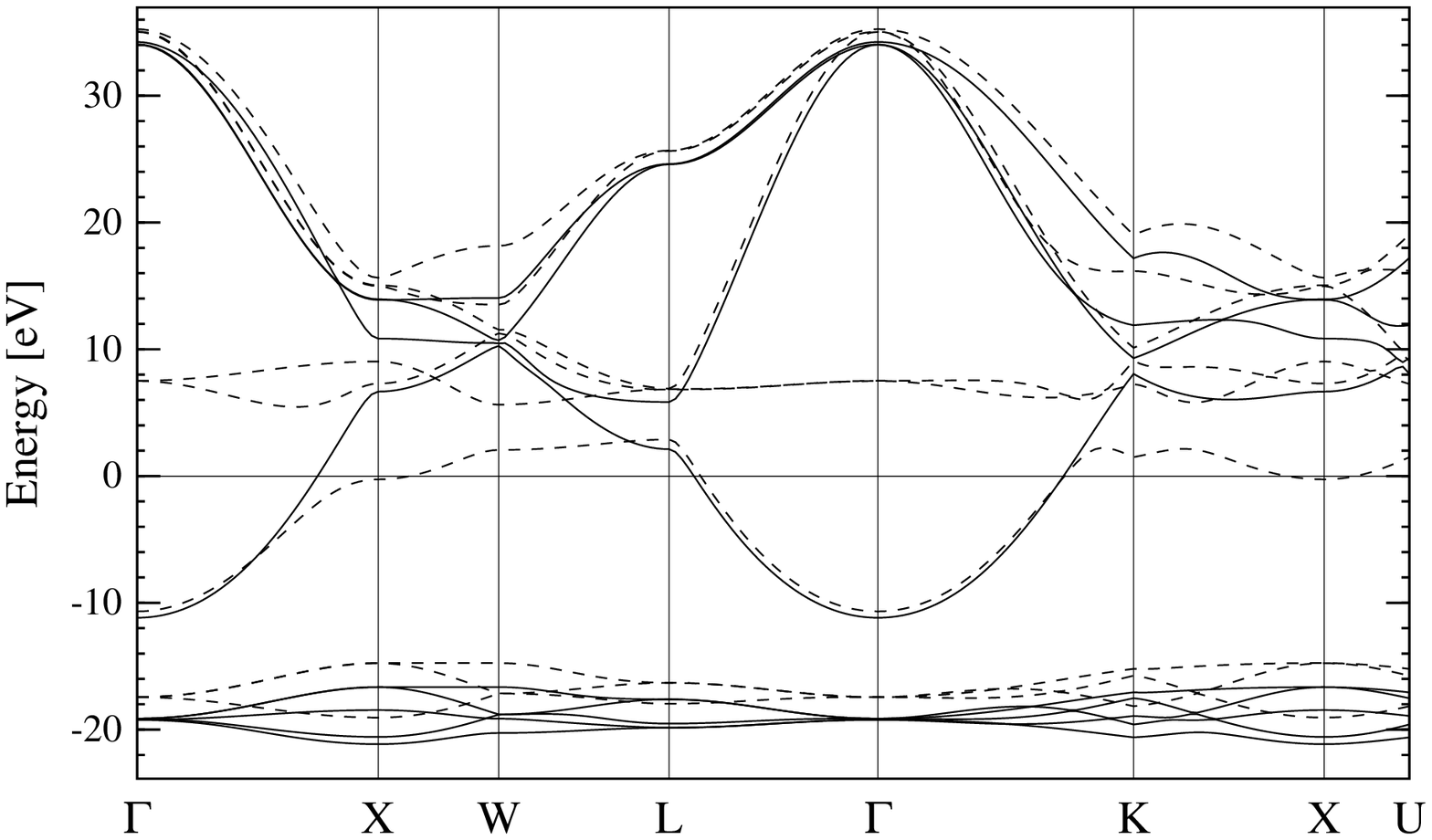}
\caption{\label{fig:cohfband} The same as Fig. \ref{fig:fehfband} for Co.}
\end{figure}

\begin{figure}
\includegraphics[scale=0.45]{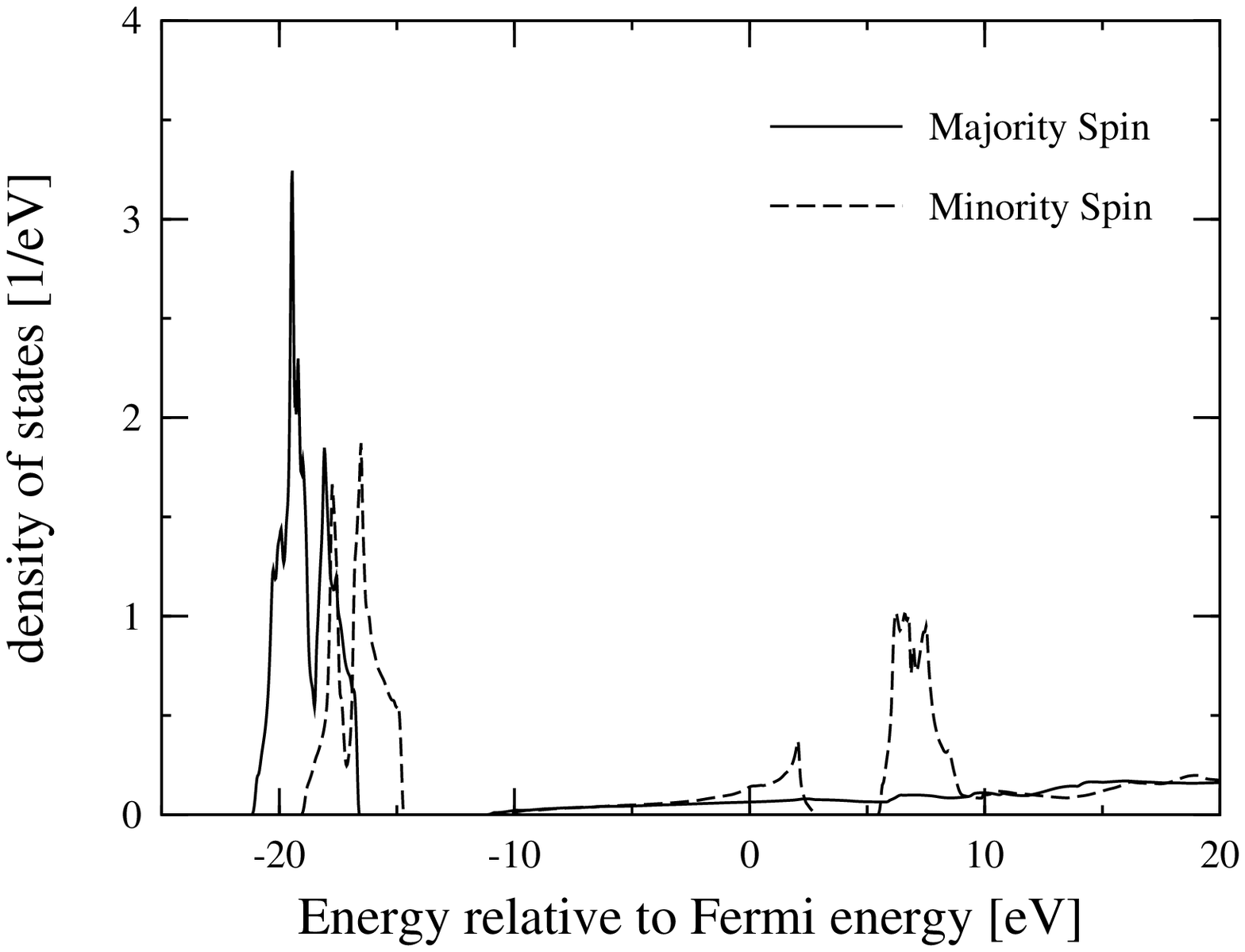}
\caption{\label{fig:cohfdos} The same as Fig. \ref{fig:fehfdos} for Co.}
\end{figure}

\begin{figure}
\includegraphics[scale=0.42]{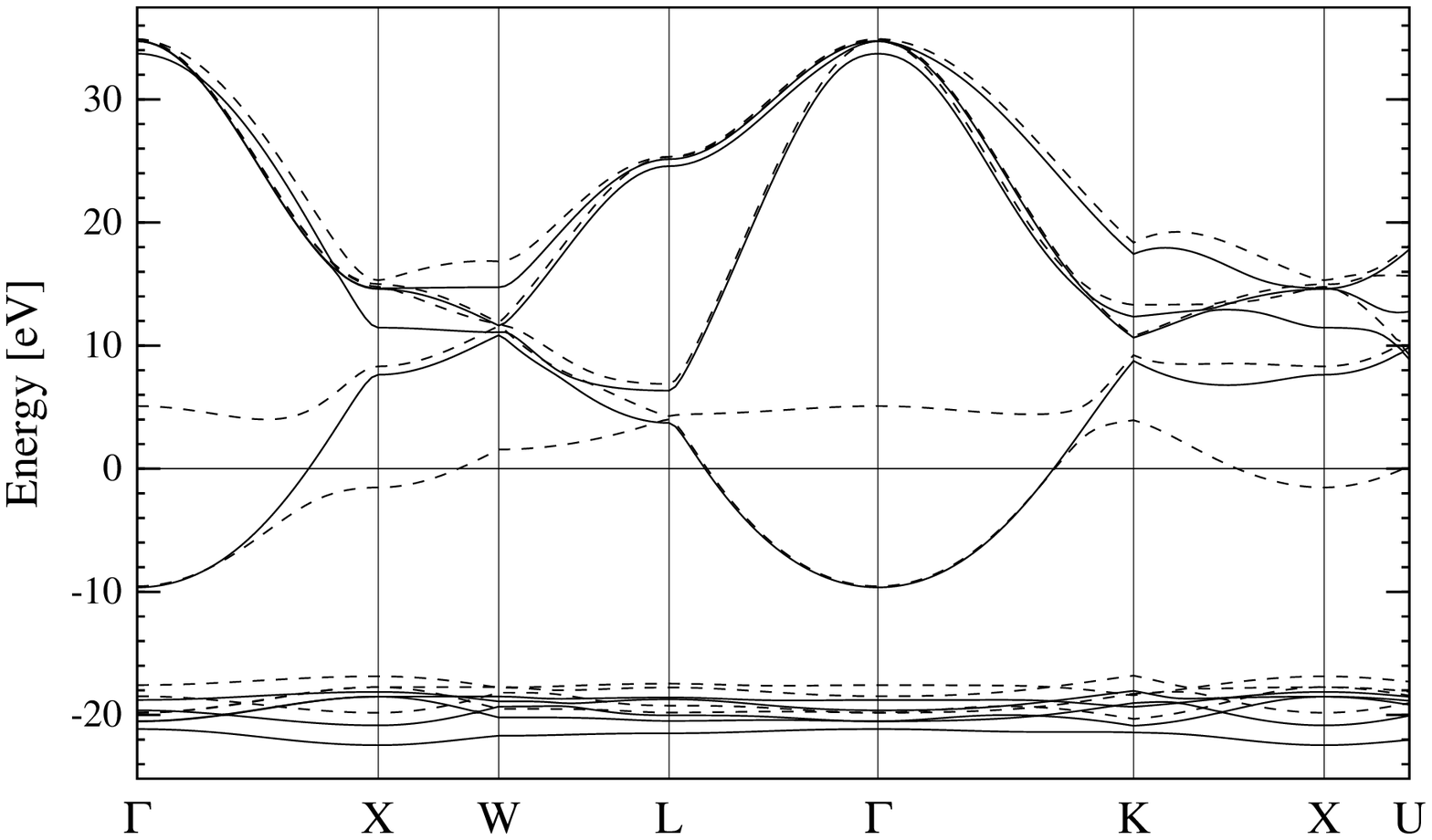}
\caption{\label{fig:nihfband} The same as Fig. \ref{fig:fehfband} for Ni.}
\end{figure}
 
\begin{figure}
\includegraphics[scale=0.45]{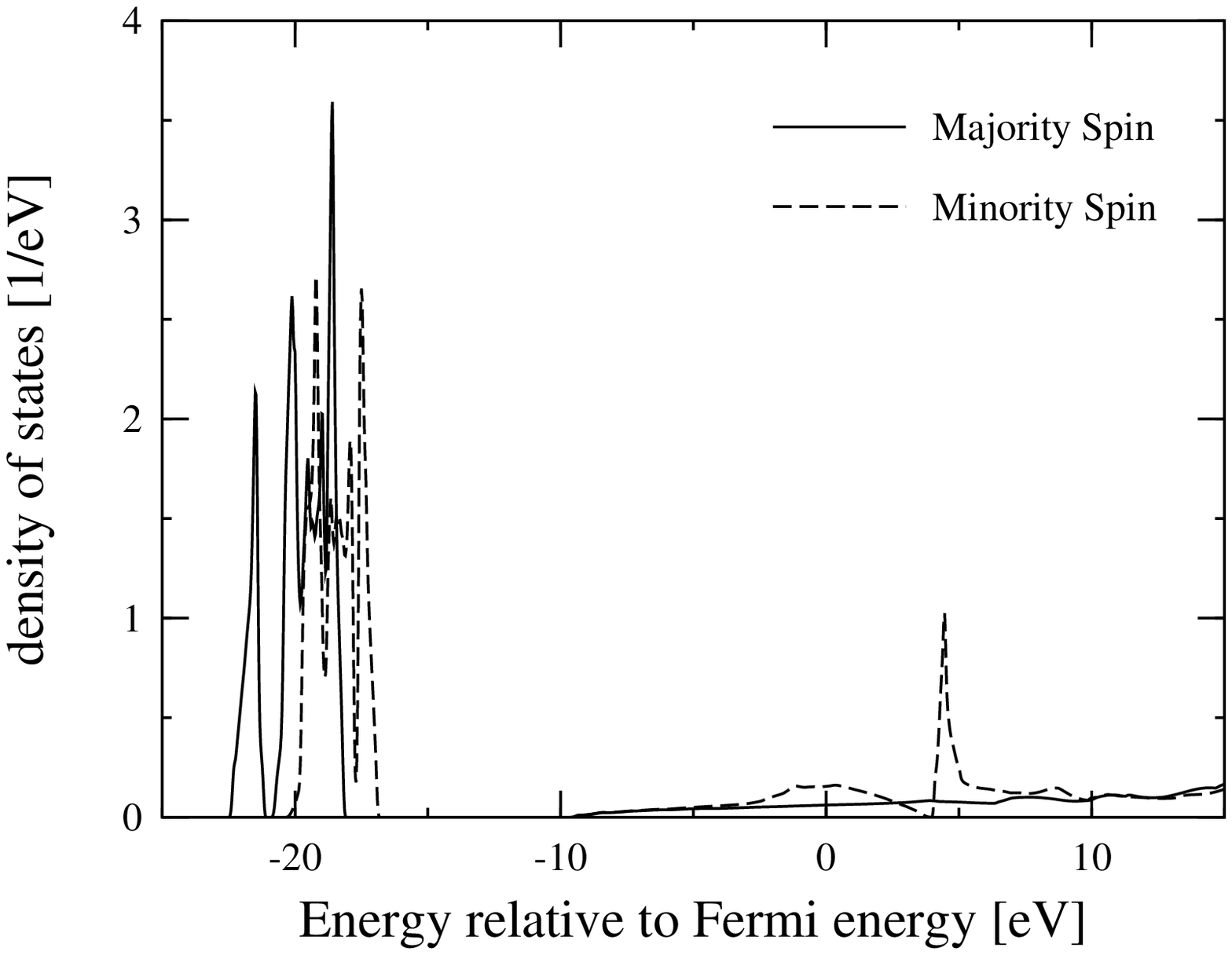}
\caption{\label{fig:nihfdos} The same as Fig. \ref{fig:fehfdos} for Ni.}
\end{figure}

\begin{figure}
\includegraphics[scale=0.42]{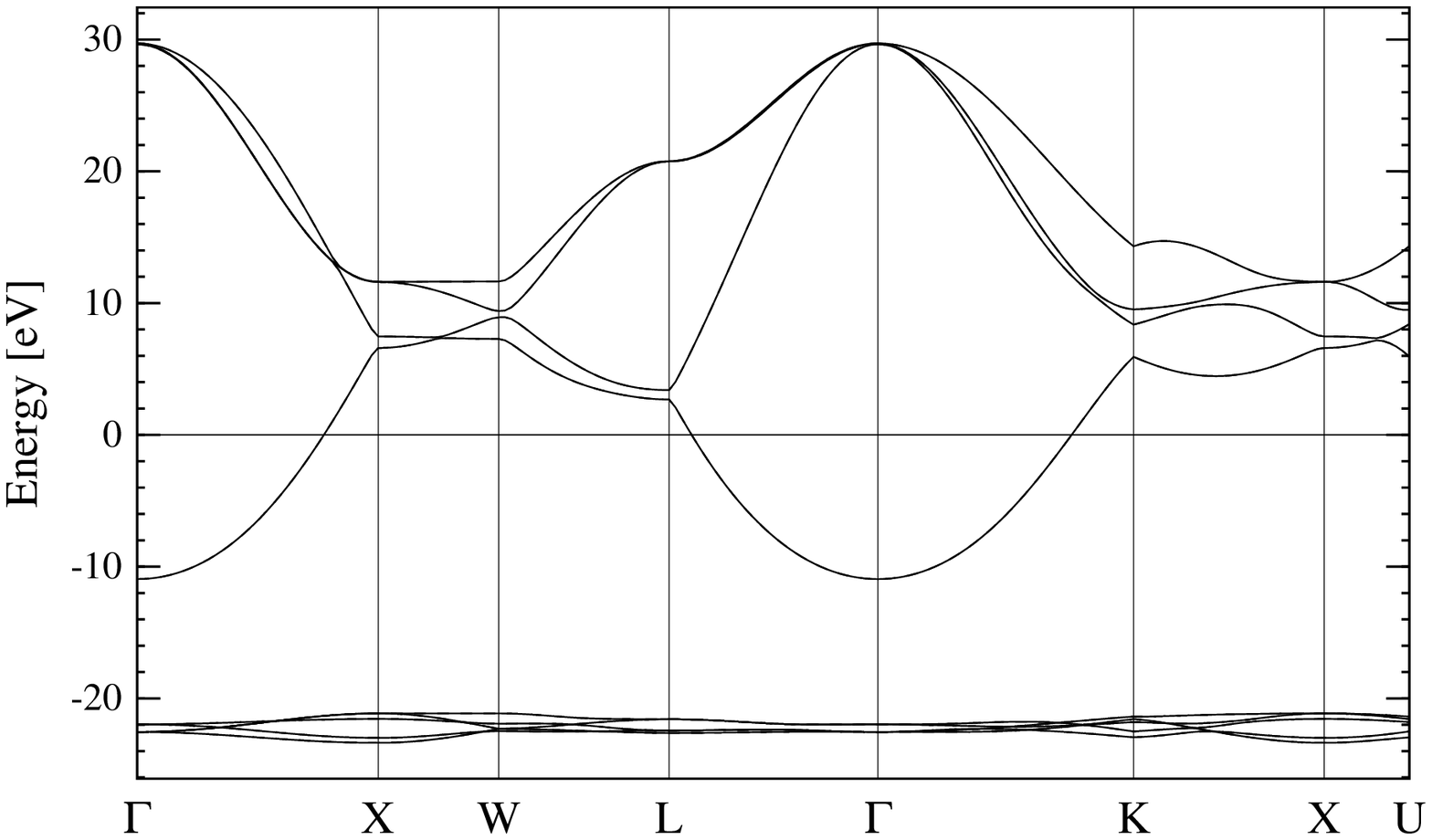}
\caption{\label{fig:cuhfband} The same as Fig. \ref{fig:fehfband} for Cu.}
\end{figure}
 
\begin{figure}
\includegraphics[scale=0.45]{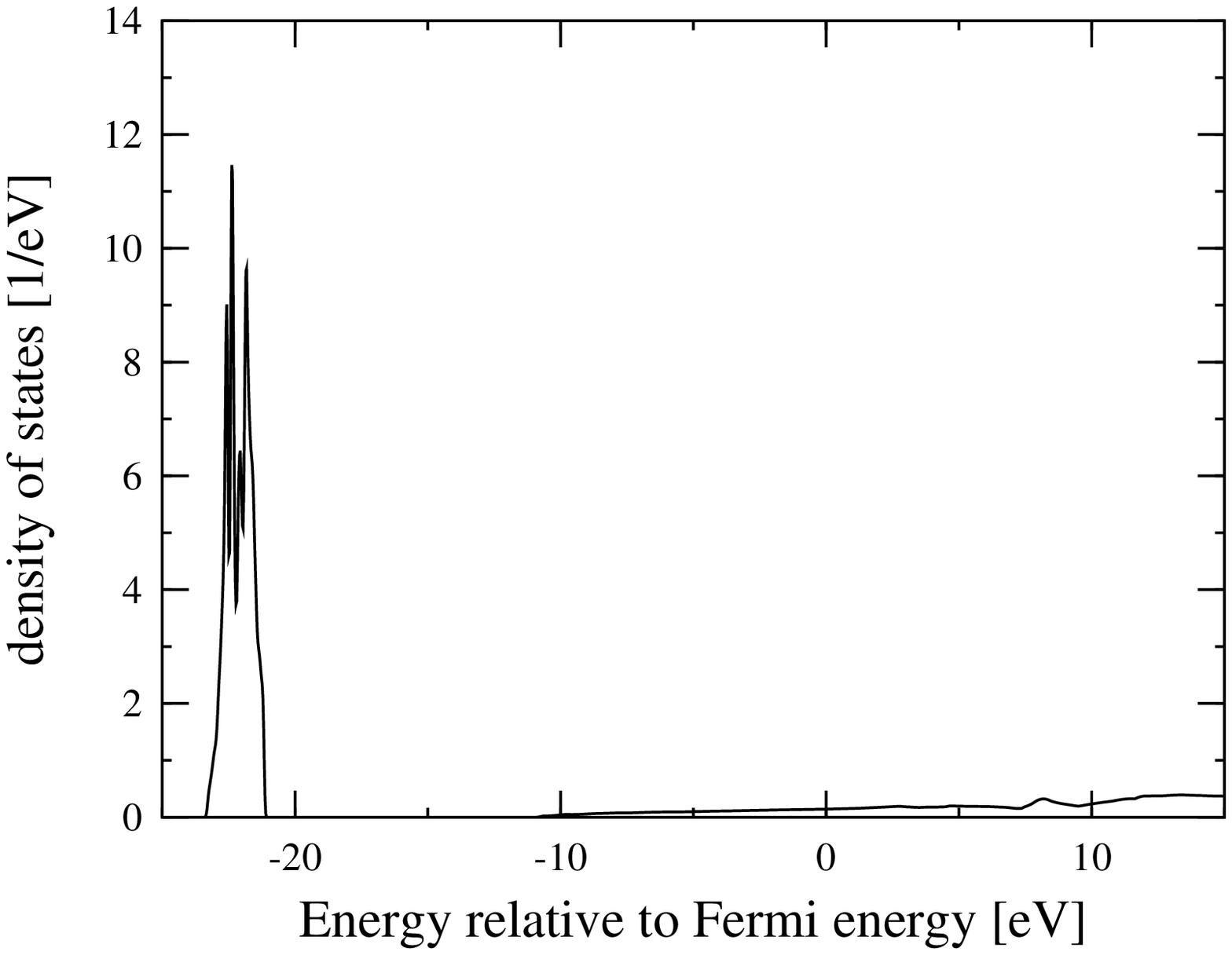}
\caption{\label{fig:cuhfdos} The same as Fig. \ref{fig:fehfdos} for Cu.}
\end{figure}

The Hartree-Fock results for the four materials of interest are shown in
Figs. \ref{fig:fehfband}--\ref{fig:cuhfdos}. We show the effective 
HF band structure and its density of states (DOS). In our HFA calculations
there are no singularities (or a vanishing DOS) at the Fermi level since
we start from a localized description and consider 
the Coulomb matrix elements only locally (on-site and for a few neighbor
shells). Therefore, we implicitely truncate the Coulomb interaction in real
space and in practice work with an effective short-ranged interaction.
Within HFA the main part of the 3d-bands lies between 18 and 22 eV
below the Fermi level and is separated from the 4sp-bands. We find
magnetism in HFA for Fe, Co, and Ni in agreement with experiment. The
five majority spin d-bands are  about 20 eV below the Fermi energy
and are completely filled. But the partially filled minority d-bands
have two (for Fe), three (for Co), and four (for Ni) filled bands between 
-18 and -15 eV, and the rest are around and above the Fermi level.
This results in magnetic moments per atom (in units of the
Bohr magneton $\mu_B$/atom) of 2.9 for iron,
1.9 for cobalt, and 0.76 for nickel.
For copper no magnetism and exchange splitting of the 3d-bands is obtained,
but the (spin degenerate) 3d-bands are at about 22 eV below the Fermi level
and separated from the 4sp-bands. If we compare these results with the
results of the simple Hartree approximation shown in
Figs. \ref{fig:feha}--\ref{fig:cuha} we see that the exchange term has two
effects:
it produces an exchange splitting and the possibility of magnetic
solutions, and it draws the 3d-bands energetically down by an amount of 
about 20 eV. Compared with experiment the HFA
overestimates magnetism and leads to overly large values for the magnetic
moment per site (about 10--30\% too large; experiment gives magnetic
moments of about 2.2 for iron, 1.7 for Co, and 0.6 for Ni).
This is consistent with Heisenberg or Ising model studies where the
mean-field approximation HFA also has the tendency to overestimate
magnetism and magnetic solutions.

\begin{figure}
\includegraphics[scale=0.42]{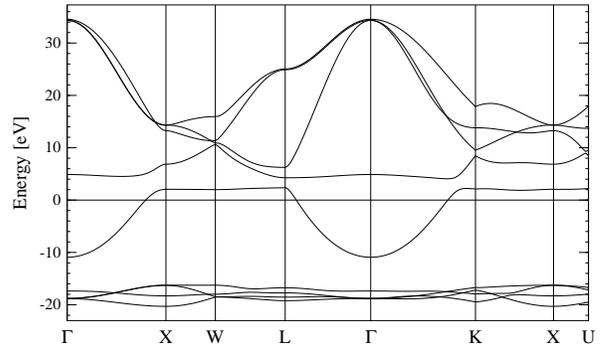}
\caption{\label{fig:conm-band} Non-magnetic HFA-band structure for Co.}
\end{figure}

\begin{figure}
\includegraphics[scale=0.45]{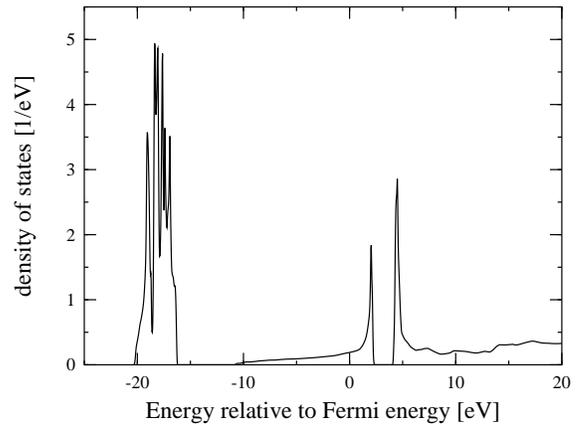}
\caption{\label{fig:conm-dos} Total density of states (of both
  degenerate spin directions) for Co obtained within the non-magnetic
  HFA-solution.}
\end{figure}

However, the reason why the 3d-bands lie so far below the Fermi level and the
4sp-band in HFA has nothing to do with the existence and overestimation of
magnetism. This can be seen already from the non-magnetic system Cu, for
which the (fully occupied) 3d-bands also lie at about 22 eV below
the Fermi level (see Figs. \ref{fig:cuhfband} and \ref{fig:cuhfdos}). 
To demonstrate this also for a system with a partially filled 3d-band we have
done a non-magnetic Hartree-Fock-calculation for Co (by forcing
equal occupation for both spin directions).  The results for the band structure
and the DOS are shown in Figs. \ref{fig:conm-band},\ref{fig:conm-dos}.
We observe again that the main part of the 3d-bands are well below the
4s-bands and Fermi level; note the hybridization gap
caused by the unoccupied 3d-bands above the Fermi level.

\section{\label{sec:atomicHFA} Comparison with atomic Hartree-Fock results}

We have seen in the previous section that one effect of the HFA calculation,
when compared with the Hartree calculation, is the shift of the 3d-bands
down (about 20 eV below the Fermi level and about 8--10 eV below the
bottom of the 4sp-band). This shift of the d-bands is about the same energy as
the Coulomb matrix elements $U$, and roughly agrees earlier atomic Hartree-Fock
calculations\cite{Watson59,HodgesWatson72}, where the 3d-states are also
about 10 eV below the 4s-states. 

\begin{figure}
\includegraphics[scale=0.45]{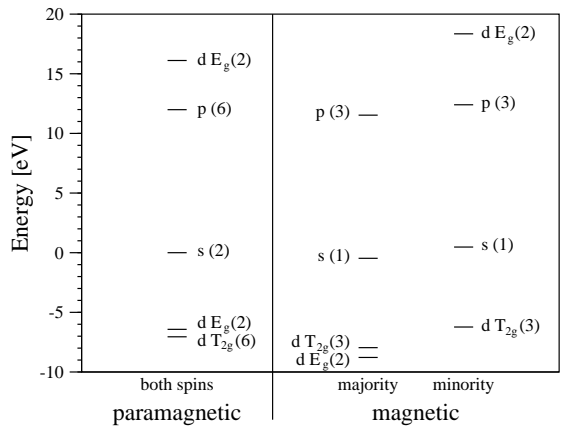}
\caption{\label{fig:coatom} Energy eigenvalues from quasi-atomic
  HFA-calculation.  The numbers in brackets indicate the degeneracy.}
\end{figure}

Because the inter-site hopping matrix elements in Table \ref{tab:UJtaverage}
are much smaller than the U-values one may consider an expansion in
$t/U$, with the zeroth order approximation to completely neglect hopping.
Doing this, we have performed a quasi-atomic HFA calculation for Co,
by including only the on-site one-particle and two-particle (Coulomb) 
matrix elements. The results are summarized in Fig. \ref{fig:coatom};
the degeneracy of the different levels is also indicated. In the paramagnetic
case,we find that the 3d-bands are below the 4s-bands (at the Fermi level) by 
about 6 to 7 eV, which is in rough agreement with the earlier atomic
HFA results\cite{Watson59,HodgesWatson72}. The splitting between the
occupied and unoccupied 3d-states is about 23 eV, which is the on-site
$U$ for Co. Magnetic HFA solutions are also found in the atomic
limit for Co, as shown in the right panel of Fig. \ref{fig:coatom}. 
The majority-spin 3d states ($T_{2g}$ and $E_g$) are now completely filled
and energetically lie lower than the corresponding non-magnetic
HFA-states. But only the (3-fold degenerate) $T_{2g}$-states of the
minority-spin electrons are filled whereas the $E_g$-states of the minority
electrons are empty (and now even 26 eV above the occupied d-states).
The additional energetical shifts between the occupied 3d-states in the
paramagnetic and ferromagnetic atomic HFA solution are due to the exchange
matrix elements $J$.

This behavior can qualitatively be understood within the framework of the
following simple, analytically solvable model.  Similar to
the numerical HFA-results presented and discused above, we neglect
all intersite one-particle (hopping) and interaction matrix elements.
Furthermore, we assume that we have diagonalized the one-particle Hamiltonian, 
taking into account only the atomic 
3d-levels and assuming that the on-site one-particle diagonal matrix elements
$\varepsilon$, the Coulomb matrix elements $U$, and the
exchange matrix elements $J$ are equal, i.e., that the 3d-levels are
degenerate in the atomic limit with no crystal-field effects. Then the
atomic part of the many-body Hamiltonian can be
written as
\begin{eqnarray} \nonumber
  H = \sum_{i\sigma} \varepsilon ~ c_{i\sigma}^{\dagger} c_{i\sigma} &+&
  \frac{U}{2} \sum_{(i\sigma) \neq (j\sigma')}
    c_{i\sigma}^{\dagger} c_{i\sigma} c_{j\sigma'}^{\dagger} c_{j\sigma'}
\\
  &+& \frac{J}{2} \sum_{i \neq j, \sigma \sigma'}
    c_{i\sigma}^{\dagger} c_{j\sigma'}^{\dagger} c_{i\sigma'} c_{j\sigma}
\end{eqnarray} 
where $i,j \in \{0,\ldots,4\}$ denote the 5 (degenerate) 3d-states.

The standard Hartree-Fock decoupling leads to
\begin{eqnarray} \nonumber
  H = \sum_{i\sigma} \Big( \varepsilon &+& U \big[ \sum_{j\sigma'}
  \langle c_{j\sigma'}^{\dagger} c_{j\sigma'} \rangle 
  - \langle c_{i\sigma}^{\dagger} c_{i\sigma} \rangle \big] 
\\
  &-& J \sum_{j \neq i} \langle c_{j\sigma}^{\dagger} c_{j\sigma} \rangle
  \Big)
  c_{i\sigma}^{\dagger} c_{i\sigma} ~~.
\end{eqnarray}
Here we have assumed that the Hartree-Fock Hamiltonian has the same symmetry 
as the uncorrelated Hamiltonian, and hence off-diagonal expectation values
$\langle c_{j\sigma'}^{\dagger} c_{i\sigma}\rangle$ for $(i\sigma) \neq (j\sigma')$
vanish. 
From this equation it is clear that the HF Hamiltonian can be written
in terms of an effective one-particle energy
\begin{equation}
 H = \sum_{i\sigma} \varepsilon_{i\sigma}^{\text{HFA}}
       c_{i\sigma}^{\dagger} c_{i\sigma} 
\end{equation}
where
\begin{equation}
  \varepsilon_{i\sigma}^{\text{HFA}} = 
  \varepsilon + U \big[
    \sum_{j\sigma'}
      \langle c_{j\sigma'}^{\dagger} c_{j\sigma'} \rangle 
    - \langle c_{i\sigma}^{\dagger} c_{i\sigma} \rangle
  \big] 
  - J \sum_{j \neq i} \langle c_{j\sigma}^{\dagger} c_{j\sigma} \rangle .
\end{equation}
In the simple Hartree approximation (HA) the exchange decouplings are
neglected, which means that all the decoupling terms with the negative
sign would not occur. Therefore, the corresponding Hartree one-particle
energies are given by
\begin{equation}
  \varepsilon_{i\sigma}^{\text{HA}} = 
  \varepsilon + U \sum_{j\sigma'}
  \langle c_{j\sigma'}^{\dagger} c_{j\sigma'} \rangle ~~.
\end{equation}
Comparing this result with the Hartree-Fock one-particle energies, we find
that the HF occupied levels are shifted downwards by an amount of
\begin{equation}
  U \langle c_{i\sigma}^{\dagger} c_{i\sigma} \rangle +
  J \sum_{j \neq i} \langle c_{j\sigma}^{\dagger} c_{j\sigma} \rangle
\end{equation}
relative to the Hartree levels.
Momentarily setting $J=0$, we see that for $N$
occupied levels the Hartree approximation gives the one-particle energies
\begin{equation}
  \varepsilon_{i\sigma}^{\text{HA}} = \varepsilon + N U 
\end{equation}
whereas the HFA yields
\begin{equation}
  \varepsilon_{i\sigma}^{\text{HFA}} = \varepsilon + (N-1) U ~~.
\end{equation}
The occupied Hartree-Fock one-particle energies are lower than the 
corresponding Hartree one-particle energies by $U$, which
is a consequence of the artificial and unphysical self-interaction still
present in the Hartree approximation that is exactly cancelled in
Hartree-Fock. This also explains why the 
Hartree-Fock bands are shifted downwards from the Hartree bands by 
an energy of the amount $U$. One also sees from this simple atomic-limit
Hartree-Fock model that the energy difference between the highest
occupied and the lowest unoccupied effective Hartree-Fock one-particle
energies is again essentially $U$, which is once more in agreement with our
numerical results for the crystal and for the atom (cf. Fig.\ref{fig:coatom}).
Note that we have ignored $U_{sd}$ interactions, which cause an additional
shift of d-bands below the s-bands by about an additional 10 eV in
the full HFA calculations.

Taking into account the exchange interaction $J$ again and denoting by
$N_{\sigma}$ the number of occupied states with
spin $\sigma$ (i.e. $N = N_{\uparrow} + N_{\downarrow}$) one obtains in HFA
\begin{equation}
  \varepsilon_{\sigma}^{\text{HFA}} =
  \varepsilon + (N-1) U - (N_{\sigma}-1) J ~~.
\end{equation}
Then the total energy in HFA is given by
\begin{equation}
  E_{\text{tot}} = N \varepsilon + \frac{N(N-1)}{2} U -
  \sum_{\sigma} \frac{N_{\sigma}(N_{\sigma}-1)}{2} J ~~.
\end{equation}
For the total energy we have added the necessary correction term to the
sum of the occupied energy levels (much like the double counting
term that shows up in band-structure calculations).
Now for partially filled 3d-shells the occupation of the different spin
directions may be different. Denoting $M = N_{\uparrow} - N_{\downarrow}$ we
obtain for the total energy
\begin{equation}
  E_{\text{tot}} = N \varepsilon + \frac{N(N-1)}{2} U - \frac{N^2+M^2}{4} J
  + \frac{N}{2} J ~~.
\end{equation}
The magnetic ($M \neq 0$) total energy is lower than the nonmagnetic
(consistent with Hund's rules). 

Take once more Co with 8 3d-electrons.  The paramagnetic
(nonmagnetic) state has the occupations
$N_{\downarrow} = N_{\uparrow} = 4$ ($N$=8 and $M$=0). 
For this configuration (corresponding to the left panel in
Fig. \ref{fig:coatom}) one obtains
\begin{equation}
  E_{\text{tot}}^{(P)} = 8 \varepsilon + 28 U -  12 J ~~.
\end{equation}
The Hund's rule magnetic solution has 3d-states of
one spin-direction completely filled, i.e.,
$N_{\uparrow} = 5$ and $N_{\downarrow} = 3$
($N$=8 and $M$=2). 
This gives
\begin{equation}
  E_{\text{tot}}^{(M)} = 8 \varepsilon + 28 U -  13 J ~~.
\end{equation}
Therefore, the magnetic configuation (with a magnetic
moment of 2 for the atom) is energetically more favorable by  $J$.
Note also the exchange splitting in the occupied energy eigenvalues
\begin{equation}
  \varepsilon_{\downarrow} - \varepsilon_{\uparrow} = 2J
\end{equation}
and that our model would predict the unoccupied minority spin $E_{2g}$ state
to be $U+J$ higher in energy than the corresponding occupied majority spin
state.

The simple model in this section differs from the results shown in
Fig.~\ref{fig:coatom} in that we have replaced the full matrix of $U$ and
$J$ by scalar values for d-states only
(ignoring s-d interactions, for example).
However, it captures all of the important physics without attempting to
be completely quantitative.

\section{\label{sec:LSDA} Comparison with LSDA results}

\begin{figure}
\includegraphics[scale=0.45]{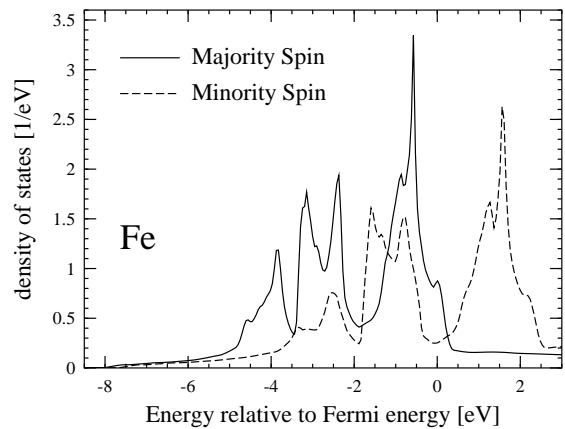}
\caption{\label{fig:felsda} Density of states (per spin direction) for
  Fe within the (unscreened) local spin-density approximation(LSDA);
  the full line shows the majority (spin up),
  the dashed line the minority spin contribution.}
\end{figure}

\begin{figure}
\includegraphics[scale=0.45]{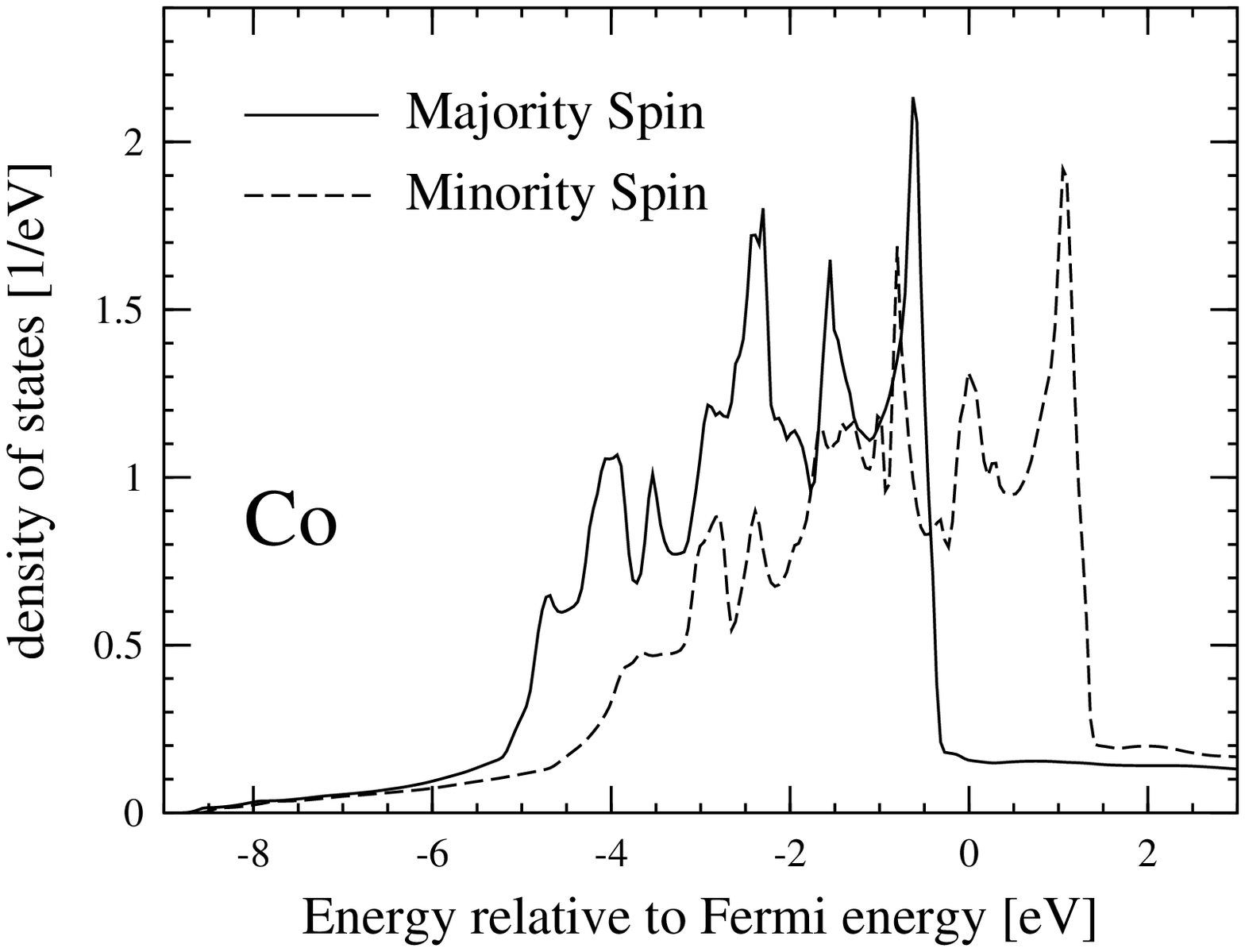}
\caption{\label{fig:colsda} The same as Fig. \ref{fig:felsda} for Co.}
\end{figure}

\begin{figure}
\includegraphics[scale=0.45]{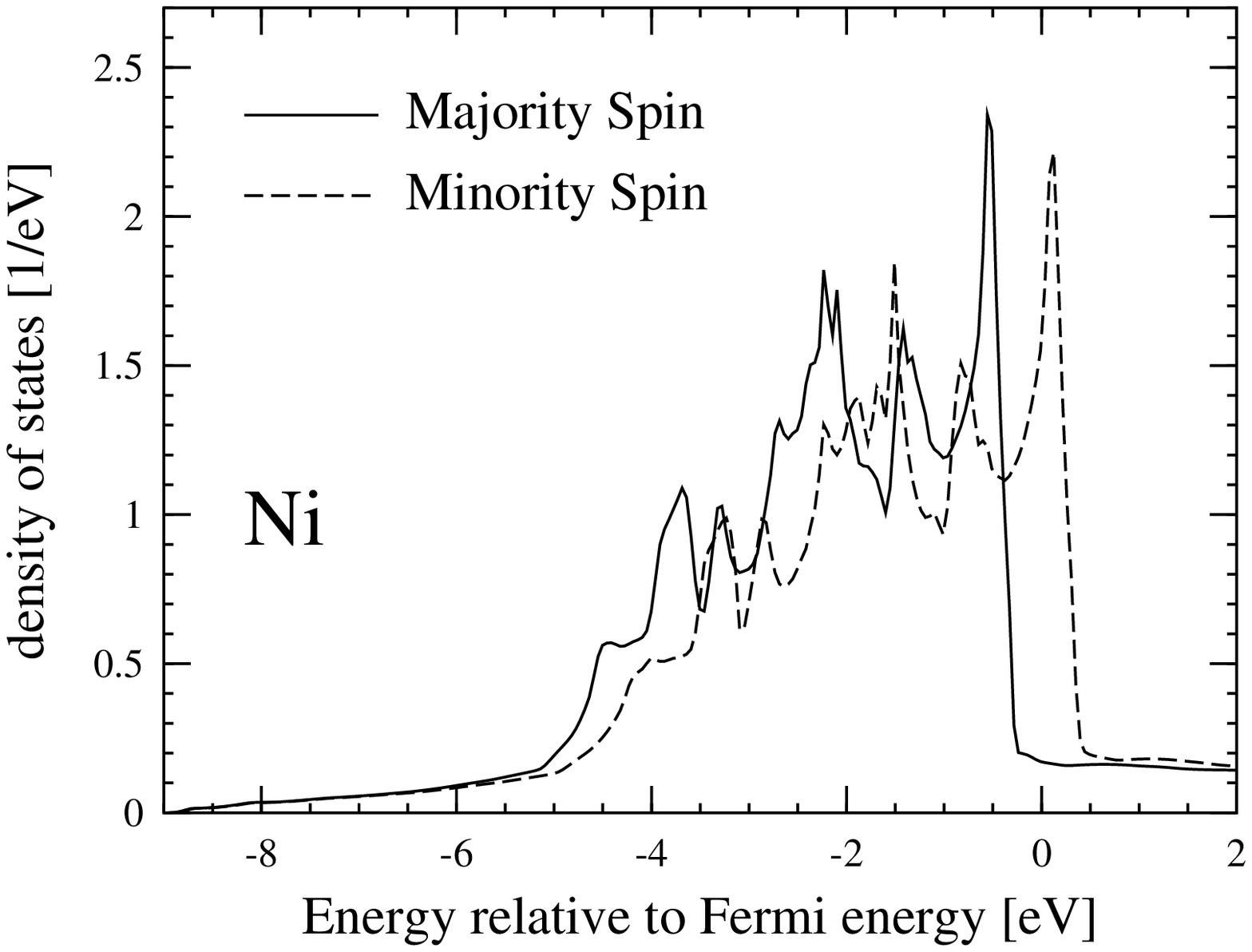}
\caption{\label{fig:nilsda} The same as Fig. \ref{fig:felsda} for Ni.}
\end{figure}

\begin{figure}
\includegraphics[scale=0.45]{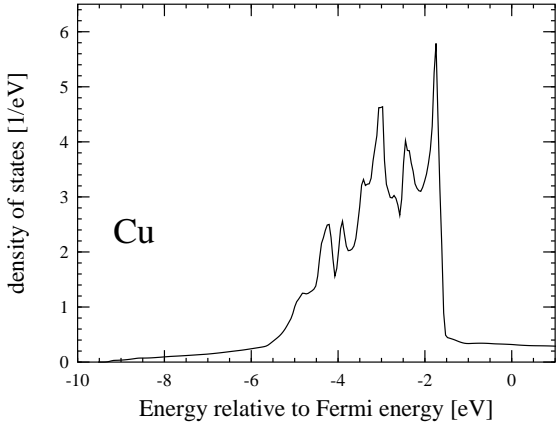}
\caption{\label{fig:culsda} Total density of states (of both degenerate
  spin directions) for Cu within the local density approximation(LSDA)}
\end{figure}

For comparison with the HFA results described in Section
\ref{sec:HartreeFock} we have also performed a standard 
local spin-density approximation (LSDA),
ab-initio, band-structure calculation with the LMTO-ASA method.
The results for the DOS are shown in Figs.
\ref{fig:felsda}--\ref{fig:culsda}. For the magnetic systems Fe, Co, and Ni
one obtains also an exchange splitting and the prediction of magnetic
solutions with magnetic moments of 2.18 for iron, 1.58 for Co, and 0.58 for
Ni, which are in better agreement with experiment than the HFA
results. This is consistent with the expectation that L(S)DA is tuned
to accurately predict ground-state properties.

The energy spectra of the bands (DOS) are quite different from the HFA.
For example, the 3d-bands now fall into the same energy region as
the 4sp-bands, i.e., the LSDA-results are not so different from the
Hartree-results. This means that the exchange-correlation energy leads only
to a small shift of the 3d-bands downwards by at most a few eV and a smaller
exchange splitting (also of the magnitude of 1 eV)
in the case of magnetic solutions. On the other hand, in the LSDA
calculations the self-interaction terms are not completely canceled, i.e.,
an (unrealistic) self-interaction is included, which may lead to 3d bands
that lie energetically too high, as discussed for the atomic limit in the
previous section.

\begin{figure}
\includegraphics[scale=0.45]{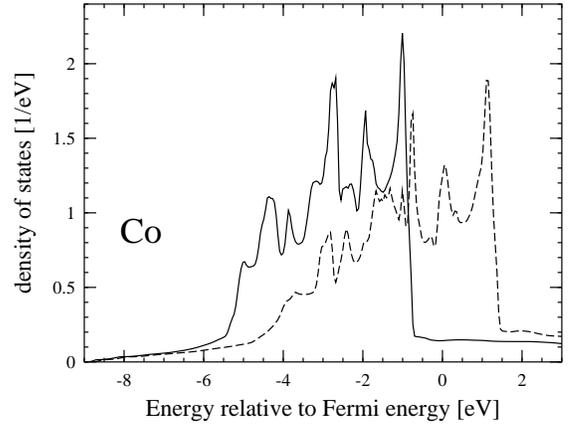}
\caption{\label{fig:coxolsda} Total density of states for Co 
  within the ''exchange-only`` local density approximation(LSDA).}
\end{figure}

To see the effect of correlations within LSDA, we have also
performed an exchange-only calculation for Co.
The result is shown in Fig.~\ref{fig:coxolsda}.
Obviously the (majority) d-bands lie lower than the
ones in the full LSDA-calculation shown in Fig.~\ref{fig:colsda}.
But the shift is of the magnitude of 1 eV only,
i.e., very minimal when compared with the large drop in the full HFA.
On the other hand, this
exchange-only LSDA-result also does not contain self-interaction corrections,
which are responsible for the large shift downwards of the d-bands. 
Nevertheless, the LSDA result indicates that a possible 
effect of correlations is to shift the 3d-bands up relative to exchange-only
calculations, and hence one would expect a similar effect if correlations
could be added to the full HFA-calculations. 

\begin{figure}
\includegraphics[scale=0.45]{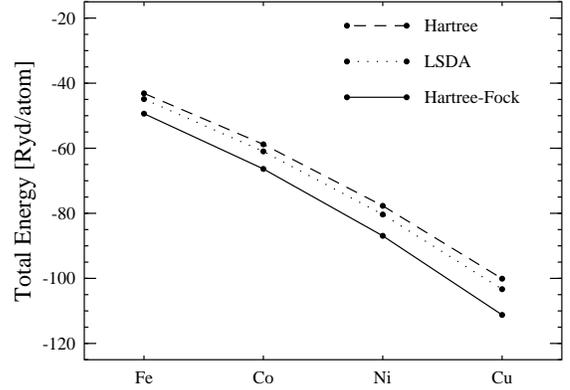}
\caption{\label{fig:etot} Total ground state energy (of the valence electrons) 
  obtained in Hartree-approximation, L(S)DA and HFA for the 3d transition
  metals Fe through Cu.}
\end{figure}

One can also calculate the total energy in the Hartree, HFA, and LSDA
approximations. The results obtained for the four materials of interest are
shown in Fig. \ref{fig:etot}. We see that the total energy is always
significantly 
lower in HFA than in the Hartree approximation, which is expected because
the HFA minimizes the total energy.  The HFA total energy is also
lower than the L(S)DA, and the LSDA result is lower than the
simple Hartree result. Because of the unknown approximations that go
into constructing L(S)DA, it is hard to guess ahead of time that this would be
the case.  However, it is well known that the L(S)DA approximation produces a
bad total exchange-correlation energy; the reason why such good agreement
with experiment is found is that relative exchange-correlation energies are
nonetheless reasonably accurately calculated.

\section{\label{discussion} Discussion and conclusion}

In this paper we have presented the results of (unscreened) HFA calculations 
for the 3d transition metals Fe, Co, Ni and Cu. We obtain magnetic solutions
for Fe, Co, and Ni with (slightly) too large magnetic moments when compared
to experimental or LSDA results. The occupied HFA 3d-bands lie about 20 eV 
below the Fermi level (and the Hartree result), which is also the magnitude of
the splitting between occupied and unoccupied 3d-bands and of the magnitude of
the on-site Coulomb matrix element (the ``Hubbard'' U). This downwards shift
of the HFA 3d-bands compared to the Hartree- and L(S)DA-3d-bands can be
understood as due to the self-interaction correction of HFA. 

One may argue that these results are not surprising and an artifact of
using the unscreened HFA. Our ab-initio calculation of
the direct Coulomb matrix elements yields large values of the magnitude of 
20 eV. Therefore the Coulomb energies are large for these materials and HFA
can be considered to be an approximation for the selfenergy which is correct
only in linear order in the Coulomb interaction. But for these large values
of the U-terms HFA is certainly not sufficient but one has to apply better
many-body approximations. One should apply systematic extensions of HFA,
which within the standard perturbational approach can be
represented by (a resummation of an infinite series of) Feynman diagrams, or
one can try to apply the recently so successful non-perturbational many-body
schemes like ``dynamical mean field theory'' (DMFT)\cite{GKKR96} or
variational (Gutzwiller) approaches\cite{GebhardWeber}. 
The simplest standard diagram series
are the bubble diagrams leading essentially to the ''random phase
approximation'' (RPA). This means just a renormalization of the interaction
line, i.e. the pure ``naked'' Coulomb interaction has to be replaced by a
``dressed'' interaction. Or in other words, the exchange (Fock) contribution
has not to be calculated with the bare Coulomb matrix elements but with
screened Coulomb matrix elements. Probably the non-perturbational schemes
like DMFT are also only applicable for screened Coulomb matrix elements.

To summarize, the new many-body ab-initio approach proposed in
Ref. \onlinecite{Schnell2003-1} is applicable to real
materials. Maximally localized Wannier functions and their one- and
two-particle matrix elements can be calculated from first principles so as
to obtain a many-body multi-band Hamiltonian in second quantization.
Standard methods of many-body theory can then be applied to this Hamiltonian.
The simplest approximation (HFA) is unreliable, because the large Coulomb
matrix elements are unscreened. Therefore, in the future better many-body
approximations should be used. Within the standard Feynman diagram approach
the most straightforward next step consists in a summation of
bubble diagrams leading to a renormalized (screened) Coulomb interaction.
This requires calculating the exchange
contribution not with the bare but with a screened Coulomb interaction. 
To take into account the effects of screening one has to calculate the
charge susceptibility and the (static) dielectric constant, which can be
done within a generalized Lindhard theory, for instance. 

\begin{acknowledgments}
  This work has been supported by a grant from the Deutsche
  Forschungsgemeinschaft No. Cz/31-12-1.
  It was also partially supported by the Department of Energy
  under contract W-7405-ENG-36.  This research used resources of the
  National Energy Research Scientific Computing Center, which is
  supported by the Office of Science of the U.S. Department of Energy
  under Contract No. DE-AC03-76SF00098.
\end{acknowledgments}

\end{document}